\documentclass[journal]{IEEEtran}
\usepackage{bbm}
\usepackage{amssymb,amsmath,amsthm}
\usepackage{graphicx,epstopdf,psfrag,url,cite,xcolor,multirow,float}
\usepackage[small,bf]{caption}
\usepackage[font=footnotesize]{subfig}

\usepackage{algorithm}
\usepackage{algpseudocode}

\newtheoremstyle{mythm}
{\topsep}   
{\topsep}   
{\itshape}      
{0pt}       
{\bfseries} 
{:}         
{5pt plus 1pt minus 1pt}    
{\thmname{#1}\thmnumber{ #2}\thmnote{ (#3)}}
\theoremstyle{mythm}

\newtheorem{proposition}{Proposition}

\newcommand{\singlesize}{0.5}

\IEEEoverridecommandlockouts

\begin{document}

\title{Capitalizing Backscatter-Aided Hybrid Relay Communications with Wireless Energy Harvesting}
\author{
Shimin Gong, \textit{Member, IEEE}, Yuze Zou, Dinh Thai Hoang, \textit{Member, IEEE}, Jing Xu, \textit{Member, IEEE}, Wenqing Cheng, \textit{Member, IEEE}, and Dusit Niyato, \textit{Fellow, IEEE}
\thanks{S.~Gong is with School of Intelligent Systems Engineering, Sun Yat-sen University, China. E-mail: gongshm5@mail.sysu.edu.cn.}
\thanks{Y.~Zou, J.~Xu, W.~Cheng are with the School of Electronic Information and Communications, Huazhong University of Science and Technology, China. E-mail: \{zouyuze,xujing,chengwq\}@hust.edu.cn.}
\thanks{D. T.~Hoang is with the Faculty of Engineering and Information Technology, University of Technology Sydney, Australia. E-mail: hoang.dinh@uts.edu.au.}
\thanks{D.~Niyato is with School of Computer Science and Engineering, Nanyang Technological University, Singapore. E-mail: dniyato@ntu.edu.sg.}
}

\maketitle
\thispagestyle{empty}

\begin{abstract}
In this work, we employ multiple energy harvesting relays to assist information transmission from a multi-antenna hybrid access point (HAP) to a receiver. All the relays are wirelessly powered by the HAP in the power-splitting (PS) protocol. We introduce the novel concept of \emph{hybrid relay communications}, which allows each relay to switch between two radio modes, i.e., the active RF communications and the passive backscatter communications, according to its channel and energy conditions. We envision that the complement transmissions in two radio modes can be exploited to improve the overall relay performance. As such, we aim to jointly optimize the HAP's beamforming, individual relays' radio mode, the PS ratio, and the relays' collaborative beamforming to enhance the throughput performance at the receiver. The resulting formulation becomes a combinatorial and non-convex problem. Thus, we firstly propose a convex approximation to the original problem, which serves as a lower bound of the relay performance. Then, we design an iterative algorithm that decomposes the binary relay mode optimization from the other operating parameters. In the inner loop of the algorithm, we exploit the structural properties to optimize the relay performance with the fixed relay mode in the alternating optimization framework. In the outer loop, different performance metrics are derived to guide the search for a set of passive relays to further improve the relay performance. Simulation results verify that the hybrid relaying communications can achieve 20$\%$ performance improvement compared to the conventional relay communications with all active relays.
\end{abstract}


\section{Introduction}

Recently, wireless backscatter has been proposed as a new promising technology to sustain wireless communications for Internet of Things (IoT)~\cite{van18}. It is featured with extremely low power consumption by transmitting information in the passive mode via the modulation and reflection of the ambient RF signals~\cite{ambient}. In contrast, the conventional RF radios operate in the active mode, which transmits information by self-generated carrier signals relying on power-consuming active components such as power amplifier and oscillator. The architectural differences between the active and passive radios lead to complement transmission capabilities and power demands in two modes, e.g.~\cite{hoang17cr} and~\cite{ieeenetwork}. In particular, the active radios may transmit with a higher data rate and better reliability, however requiring more power consumption. While the passive radios generally achieve a lower data rate with significantly reduced power consumption. This implies that a hybrid wireless system with both active and passive radios can flexible schedule the data transmissions in different modes according to the channel conditions, energy status, and traffic demands, and thus achieve a higher network performance, especially for energy constrained IoT networks.


Due to the extremely low power consumption of the passive radios, it becomes promising to use the passive relays to assist the active RF communications, especially for wirelessly powered IoT networks with energy harvesting constraints, e.g.,~\cite{lxrelay19,tccn19,R7}. However, most of the existing works have considered a simple passive relay model in which one or more backscatter radios are employed to assist the active RF communications. In this paper, we propose a novel \emph{hybrid relay communications} model in which both the passive and active relays are employed simultaneously to assist the active RF communications. Based on the dual-mode radio architecture~\cite{ieeenetwork}, each relay can switch between the active and passive modes independently to maximize the overall throughput, according to its channel conditions and energy status. With more relays in the passive mode, less active relays can be used for amplifying and forwarding the source signal. On the other hand, the active relays' channels can be enhanced by the passive relays. It is obvious that the coupling between relays in two modes complicates the optimization of individual relays' mode selection. As the conventional relay models focus on either the active or passive relays, the current relay strategies are not applicable to hybrid relay communications, which motivates our novel algorithm design in this paper.

Specifically, we focus on a two-hop hybrid relay communication model in this paper. In the first hop, the multi-antenna transmitter beamforms information to the relays and the receiver. The set of passive relays instantly backscatter the RF signals to enhance signal reception at both of the active relays and the receiver. In the second hop, the set of active relays jointly beamform the received signals to the receiver. Meanwhile, the passive relays can adapt their reflection coefficients to enhance the active relays' forwarding channels. We aim to maximize the overall network throughput by jointly optimizing the transmit beamforming, the relays' mode selection, and their operating parameters. It is clear that the throughput maximization problem is combinatorial and difficult to solve optimally. To overcome this difficulty, we propose a two-step solution to optimize the relay strategy. A preliminary study on hybrid relay communications has been reported in our previous work~\cite{xiegc19}. In this paper, we provide different evaluations of the relay performance, which motivate us to design a set of iterative algorithms that can be more effective than that in~\cite{xiegc19}. Specifically, with a fixed relay mode, we firstly find two feasible lower bounds on the signal-to-noise ratio (SNR) at the receiver under different channel conditions, based on which we devise a set of performance metrics to evaluate individual relay's performance gain. Then, we provide an iterative procedure to update the relays' mode selection that can improve the overall relay performance in each iteration. Simulation results verify that the proposed hybrid relay strategy can significantly improve the throughput performance compared to the conventional relay strategy with all active relays.

To be specific, our main contributions in this paper are summarized as follows:
\begin{enumerate}
  \item A novel hybrid relay communications model: Different from the conventional relay communications, we allow multiple energy harvesting relays in both the active and passive modes to collaborate in relay communications. The active relays follow the amplify-and-forward protocol while the passive relays can adapt their reflection coefficients to enhance the relay channels.
  \item Lower bounds on relay performance: We propose two lower bounds to evaluate the receiver's SNR under different channel conditions. The SNR evaluation serves as a performance metric for relay mode selection. Each lower bound requires to solve an optimization problem involving the transmit beamforming, the active relays' power control, and the passive relays' phase control.
  \item Heuristic algorithms for mode selection: To bypass the complexity in SNR evaluation, we also propose a set of heuristic algorithms based on simple approximations of the SNR performance to evaluate each passive relay's performance gain. Our simulation results verify that the SNR-based mode selection can achieve the optimal throughput performance, while the heuristic algorithms also perform well with significantly reduced complexity.
\end{enumerate}

The rest of this paper is organized as follows. Section~\ref{sec_related} summarizes recent works related to hybrid relay communications. Section~\ref{sec_model} describes the system model for hybrid relay communications. Correspondingly, Section~\ref{sec_prob} formulates the throughput maximization problem. In Section~\ref{sec_max}, we devise iterative algorithms to search for the optimal passive relays and the operating parameters of the active relays. Simulation and conclusions are drawn in Sections~\ref{sec_sim} and~\ref{sec_con}, respectively.

\section{Related Works}\label{sec_related}

The concept of hybrid backscatter communications has been proposed in~\cite{kim17} to increase the transmission range of wireless-powered communications networks. However, it restricts the user devices to switch between two typical configurations of backscatter communications, i.e., the bistatic and ambient configurations, e.g.,~\cite{van18} and~\cite{ieeenetwork}. The authors in~\cite{kim17} verify that an optimized time allocation between two backscatter configurations can achieve improved throughput performance and increased transmission range. Different from~\cite{kim17}, in this work we allow each user device to operate in either the active or the passive mode. By optimizing the users' radio modes, we envision to achieve enhanced network performance in terms of both energy consumption and throughput in a hybrid radio network, which have been corroborated by both analytical and simulation results in the literature. The authors in~\cite{hoang17cr} allow the cognitive radios to switch among energy harvesting, backscatter and active communications. To maximize the performance gain, the time transitions are optimized under different channel conditions according to the presence of primary users. Considering point-to-point communications between a pair of IoT devices, the authors in~\cite{luxiao18} present a theoretical analysis on the link capacities when the transmitter follows pre-designed protocols to switch between the passive and active modes. The authors in~\cite{liangaccess,lyu17,R1} allow the transmitter to switch its radio mode during data frames and further optimize its time transitions to maximize the overall throughput performance. The authors in~\cite{xulong} propose to integrate the passive backscatter communications with the cognitive radios, and verify a significant performance gain in terms of the coverage probability and ergodic capacity.

Besides mode switching of a single device, the user devices in different radio modes can further collaborate in data transmissions, exploiting both the radio diversity and user cooperation gains. The user cooperation can be envisioned by allowing the passive radios to serve as wireless relays for the active RF communications, e.g.,~\cite{ieeenetwork} and~\cite{tccn19}. Compared to the conventional relay communications, the backscatter-aided passive relay communications can be more energy- and spectrum-efficient due to the extremely low power consumption of passive radios. The authors in~\cite{yang18} show that the backscattered signals can be used to enhance the signal reception performance of the active RF communications. Different from~\cite{yang18}, the authors in~\cite{R5} and~\cite{lyu19} employ the active radios to assist information transmission of the passive radios. The rate maximization problem is formulated by jointly optimizing the time scheduling, power allocation, and energy beamforming strategies. The relay needs to decode the backscattered information in the first phase and then forward it to the receiver by active RF communications in the second phase. This implies that the relay's operation requires significant power consumption in receiving and forwarding, which may prevent it from joining the cooperative transmissions. In~\cite{Gong2018Passive}, both the passive and active radios can assist each other in a two-hop relay protocol. The source node firstly transmits information to the relay node via backscatter communications in the first hop. Then, the relay node will decode and forward the information by active RF communications, meanwhile a set of passive radios are selected and optimized to assist the active relay's RF communications in the second hop. A similar two-hop relay model is also studied in~\cite{R8}, where the relay firstly decodes the information and then uses backscatter communications to forward the information to the receiver.

The passive relay communications are also studied for multi-user wirelessly powered communications. In our previous work~\cite{tccn19}, we allow each passive relay to assist multiple active radios. Considering the couplings among different users, heuristic algorithms are devised to maximize the overall relay performance by jointly optimizing the transmit beamforming, mode switching, and the passive relays' reflection coefficients, subject to the relays' energy harvesting constraints. In~\cite{jc-iot}, the couplings among multiple energy harvesting relays are further characterized in a distributed game model, in which each passive relay aims to optimize its reflection coefficients for the other receivers to maximize its utility function.

\section{System model}\label{sec_model}

We consider a downlink communication system with a group of single-antenna user devices coordinated by a multi-antenna hybrid access point (HAP), which constantly distributes data stream to individual users. For example, we can envision the HAP as the centralized controller in a wireless sensor network for control information distribution to spatially distributed wireless sensors, which can be low-power and long-lasting IoT devices. The HAP's information transmissions to different user devices can be scheduled in orthogonal channels, e.g.,~via a time division multiple access (TDMA) protocol~\cite{tccn19}. Without loss of generality, we focus on the simplest case with only one receiver. The user devices can serve as the wireless relays for each other via device-to-device (D2D) communications. In particular, the data transmission from the HAP to the receiver can be assisted by a set of relays following the amplify-and-forward (AF) protocol. The set of relays is denoted by $\mathcal{N}=\{1,2,\ldots,N\}$. The HAP has a constant power supply, while the relays are wirelessly powered by RF signals emitted from the HAP. Via signal beamforming, the HAP can control the information rate and power transfer to the relays following the power-splitting (PS) protocol~\cite{relay13}. Our previous work in~\cite{gsmwcnc19} reveals that the relay optimization with the typical time-switching protocol has a similar structure with the PS protocol. Hence, we focus on the PS protocol in this work. Our goal is to maximize throughput from the HAP to the receiver by optimizing the HAP's transmit beamforming and the relay strategy. Assuming that the HAP has $K$ antennas, let ${\bf f}_0 \in \mathbb{C}^K$ and ${\bf f}_n \in\mathbb{C}^K$ denote the complex channels from the HAP to the receiver and from the HAP to the $n$-th relay, respectively. Let ${\bf g}\triangleq[g_1, g_2,\ldots, g_N]^T\in\mathbb{C}^N$ denote the complex channels from the relays to the receiver. We assume that all the channels are block fading and can be estimated in a training period before data transmissions~\cite{jc-iot}.

\subsection{Two-Hop Hybrid Relaying Scheme}

\begin{figure}[t]
\centering
\includegraphics[width=0.45\textwidth]{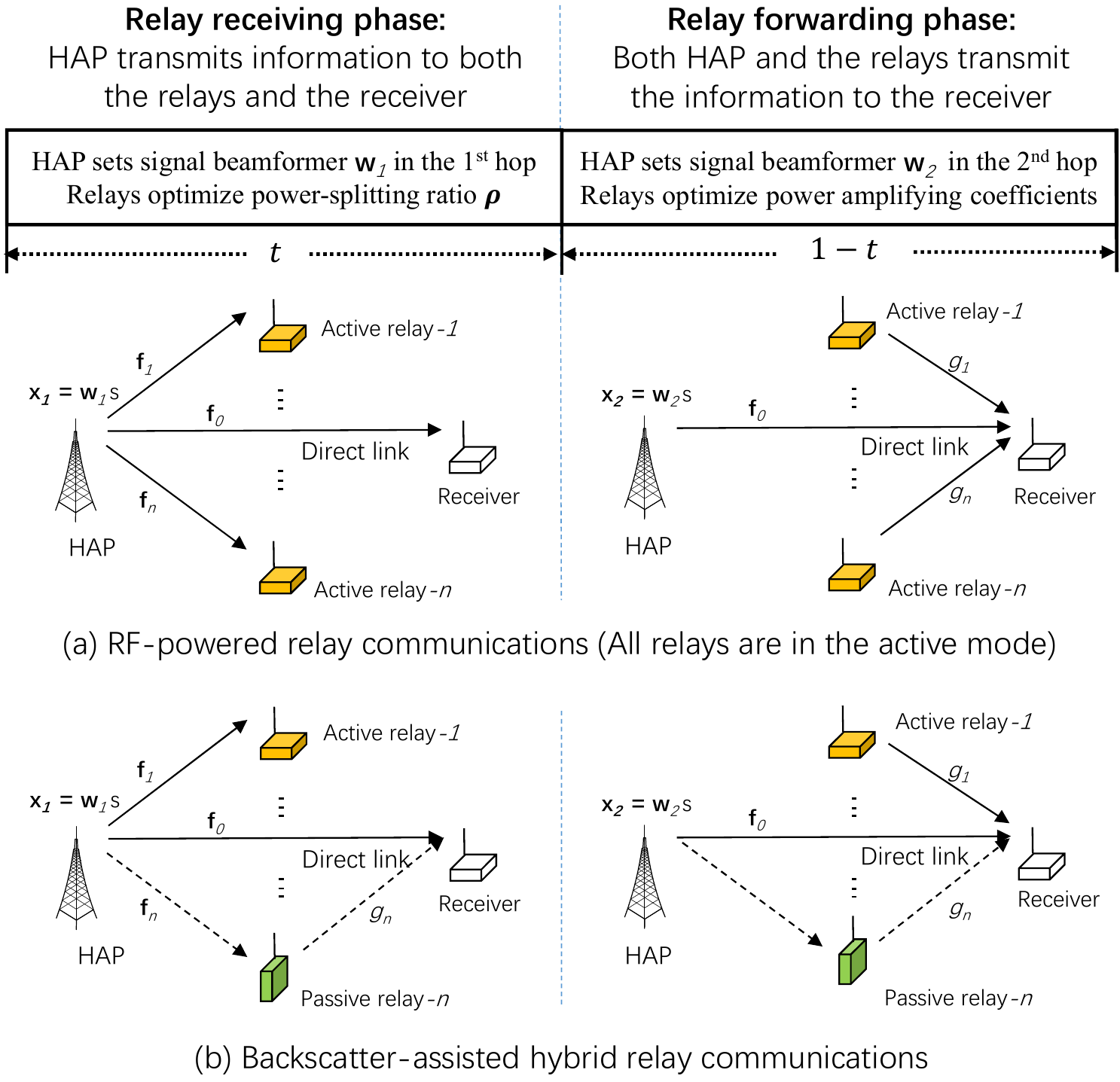}
\caption{Two-hop data transmissions in hybrid relay communications.}\label{fig_hybridmodel}
\end{figure}

The relay-assisted information transmission follows a two-hop half-duplex protocol. As shown in Fig.~\ref{fig_hybridmodel}, the information transmission is divided into two phases, i.e., the relay receiving and forwarding phases, corresponding to the information transmission in two hops. Due to a short distance between transceivers in a dense D2D network, the direct links between the HAP and the receiver can exist in both hops and contribute significantly to the overall throughput. Moreover, leveraging the direct links, we allow the HAP to beamform the same information twice in two hops. This can significantly improve the data rate or transmission reliability, which may be more crucial for many applications such as industrial control process. Specifically, let $({\bf w}_1,{\bf w}_2)$ denote the HAP's signal beamforming strategies in two phases.

In the first hop, the HAP beamforms the information with a fixed transmit power $p_t$ and the beamforming vector ${\bf w}_1$. Conventionally, the beamforming information can be received by both the relays and the receiver directly, as shown in Fig.~\ref{fig_hybridmodel}(a). Hence, the HAP's beamforming design has to balance the transmission performance to the relays and to the receiver. A higher rate on the direct link potentially degrades the signal quality at the relays and reduces the data rate of relays' transmission. Different from the conventional relay communications in Fig.~\ref{fig_hybridmodel}(a), where all relays are operating in the AF protocol, in this paper we assume that each relay has a dual-mode radio structure that can switch between the passive and active modes, similar to that in~\cite{ieeenetwork} and~\cite{jc-iot}. This arises the novel hybrid relay communications model. As illustrated in Fig.~\ref{fig_hybridmodel}(b), when the HAP beamforms the information signal to the relays, the relay-$n$ can turn into the passive mode and backscatter the RF signals from the HAP directly to the receiver. By setting a proper load impedance and thus changing the antenna's reflection coefficient~\cite{van18}, the passive relay can backscatter a part of the incident RF signals, while the other part is harvested as the power to sustain its operations. Moreover, the backscattered signals from the passive relays can be coherently added with the active relays' signals to enhance the signal strength at the receiver~\cite{yang18}.

The HAP's beamforming in the first hop is also used for wireless power transfer to the relays. We consider a PS protocol for the energy harvesting relays, i.e., a part of the RF signal at the relays is harvested as power while the other part is received as information signal. Specifically, we allow each active relay to set the different PS ratio to match the HAP's beamforming strategy and its energy demand. In the second hop, the active relays amplify and forward the received signals to the receiver. Meanwhile, the HAP also beamforms the same information symbol directly to the receiver with a new beamforming vector ${\bf w}_2$. Hence, the received signals at the receiver are a mixture of the signals forwarded by the relays and the direct beamforming from the HAP. With maximal ratio combining (MRC) at the receiver~\cite{liu2016wireless}, the received signals in both hops can be combined together to enhance the data rate and reliability in transmission. Note that the passive relays in the second hop can optimize their reflection coefficients to enhance individual relays' forwarding channels, as well as the direct channel from the HAP to the receiver.

%


\subsection{Channel Enhancement via Passive Relays}

The optimal selection of each relay's radio mode is complicated by the relays' couplings in transmission capabilities. The passive relays can enhance the active relays' channel conditions, while more active relays can achieve user cooperation gain via network beamforming~\cite{networkbf}. In the extreme case, the passive mode becomes the only afordable choice when the relay have low power supply, while the active mode can be preferred to provide more reliable transmissions if the relay has good channel conditions and sufficient power supply.

Let $b_n\in\{0,1\}$ denote the binary variable indicating the radio mode of the relay-$n$ for $n\in\mathcal{N}$. Then the set of relays in Fig.~\ref{fig_hybridmodel}(b) will be split into two subsets, i.e., $\mathcal{N}_a \triangleq \{n\in\mathcal{N}: b_n = 0\}$ and $\mathcal{N}_b \triangleq \mathcal{N}\setminus\mathcal{N}_a=\{n\in\mathcal{N}: b_n = 1\}$, denoting the sets of active and passive relays, respectively. Let $\hat{\bf f}_0$ and $\hat{\bf f}_k$ for $k\in\mathcal{N}_a$ denote the enhanced channels from the HAP to the receiver and to the active relays, respectively. Let $s(t)$ denote the signal transmitted from the HAP with a constant transmit power $p_t$. The received signal at the passive relay-$n$ is given by $y_n(t) = \sqrt{p_t}{\bf f}_ns(t)+v(t)$, where $v(t)$ is the normalized noise signal. The complex reflection coefficient of passive relay-$n$ is given by $\Gamma_n= |\Gamma_n|e^{j\theta_n}$, where $|\Gamma_n|$ denotes the magnitude of signal reflection and $\theta_n\in[0,2\pi]$ represents the phase offset incurred by backscattering. Hence, the signal received at the receiver is given by
\begin{align*}
d_r&=\sqrt{p_t}{\bf f}_0s(t)+ \sum_{n\in\mathcal{N}_b}\sqrt{p_t}{\bf f}_ns(t)\Gamma_ng_n+\bar{v}(n)\\ \notag
   &=\sqrt{p_t}\hat{\bf f}_0s(t) + \bar{v}(n)\notag,
\end{align*}
where $\bar{v}(n)$ denotes the aggregate noise signal. Hence, the enhanced channel $\hat{\bf f}_0$ from the HAP to the receiver can be rewritten as follows:
\begin{equation}\label{equ_channel_direct}
\hat{\bf f}_0={\bf f}_0 + \sum_{n\in\mathcal{N}_b}{\bf f}_n\Gamma_ng_n={\bf f}_0+\sum_{n\in\mathcal{N}}b_n{\bf f}_n\Gamma_ng_n.
\end{equation}
For single relay case, the channel model in~\eqref{equ_channel_direct} is degenerated to a simple representation as that in~\cite{yang18} and~\cite{jc-iot}. In the same way, we can rewrite the enhanced channel from the HAP to the active relay-$k$ as follows:
\begin{equation}\label{equ_channel_relay}
\hat{\bf f}_k={\bf f}_k+\sum_{n\in\mathcal{N}_b}{\bf f}_n\Gamma_nz_{n,k}={\bf f}_k+\sum_{n\in\mathcal{N}}b_n{\bf f}_n\Gamma_nz_{n,k},
\end{equation}
where $z_{n,k}$ denotes the complex channel from the passive relay-$n$ to the active relay-$k$. Note that the enhanced channels $\hat{\bf f}_0$ and $\hat{\bf f}_k$ depend not only on the binary indicator $b_n$, but also the complex reflection coefficient $\Gamma_n$ of each passive relay in the set $\mathcal{N}_b$. We observe that the phase $\theta_n$ is a critical design variable for channel enhancement while $|\Gamma_n|$ can be simply set to its maximum $\Gamma_{\max}$ to increase the reflected signal power.

The channel models in~\eqref{equ_channel_direct} and~\eqref{equ_channel_relay} are simplified approximations as we omitted the interactions among different passive relays. In fact, besides reflecting the active radios' signals, each passive relay can also reflect the backscattered signals from the other passive relays, thus creating a feedback loop, which makes it very complicated to characterize the the signal models exactly. However, this simplified model can be reasonable as the double reflections can be significantly weakened. Most importantly, it not only provides the basis for analytical study, but also enables us to derive a lower bound on the throughput performance of hybrid relay communications.


\section{Problem Formulation}\label{sec_prob}

In the sequel, we focus on this simplified channel model and formulate the throughput maximization problem for hybrid relay communications, which involves the joint optimization of the HAP's transmit beamforming, the relays' mode selection, and their operating parameters. Assuming a normalized noise power, the receiver's SNR in the first hop is given by
\begin{equation}\label{equ_gamma1}
\gamma_1 =  p_t|\hat{\bf f}_0^H {\bf w}_1|^2,
\end{equation}
where $\hat{\bf f}^H_0$ is the Hermitian transpose of $\hat {\bf f}_0$. By controlling the beamformer ${\bf w}_1$ in the first hop, the HAP can adjust its information and power transfer to different relays. Each active relay $n\in\mathcal{N}_a$ can choose the different PS ratio $\rho_n$ to balance its power supply and demand, taking into account the power budget constraint as follows:
\begin{equation}\label{equ_lineareh}
p_n \leq \eta\rho_n  p_t |\hat{\bf f}_n^H{\bf w}_1|^2, \quad \forall\, n\in\mathcal{N}_a,
\end{equation}
where $p_n$ denotes the relay's transmit power and $\eta$ is the energy harvesting efficiency. The relay's overall power consumption can also take into account a constant circuit energy consumption, which however will not bring new challenge in the following algorithm design. The PS ratio $\rho_n$ indicates the portion of RF power that is converted by the energy harvester. The other part $1-\rho_n$ is then used for signal detection and thus the received signal of the relay-$n$ is given by
\[
r_n = \sqrt{(1 - \rho_n)p_t} \hat{\bf f}_n^H {\bf w}_1 s + \sigma_n = y_n s + \sigma_n,
\]
where we define $y_n \triangleq \sqrt{(1-\rho_n)p_t} \hat{\bf f}_n^H {\bf w}_1$ for notational convenience and $\sigma_n$ is the complex Gaussian noise with zero mean and normalized unit variance. 

In the second hop, each active relay-$n$ forwards the information to the receiver with the transmit power $p_n$. All the relays' signals will be combined coherently at the receiver. The power amplifying coefficient of the relay-$n$ can be denoted as $x_n \triangleq \left(\frac{p_n}{1+|y_n|^2}\right)^{1/2}$, which has to be optimized to maximize the overall throughput~\cite{networkbf}. Meanwhile, the HAP can transmit the same information directly to the receiver with a new beamformer ${\bf w}_2$. This can enhance the reliability and data rate of information transmission from the HAP to the receiver. Hence, the received signals at the receiver is a mixture of the HAP's direct beamforming and the relays' forwarding, i.e.,
\begin{equation}\label{equ_rd}
r_d =  \sum_{n\in\mathcal{N}_a} x_n {\hat g}_n y_n s+ \sum_{n\in\mathcal{N}_a} x_n  {\hat g}_n \sigma_n + \sqrt{p_t}\hat{\bf f}_0^H {\bf w}_2 s + v_d,
\end{equation}
The first two terms in~\eqref{equ_rd} correspond to the amplified signals by the active relays. The third term represents the direct beamforming from the HAP. Due to the passive relays' backscattering, the channel $\hat g_n$ is also an enhanced version of $g_n$ from the relay-$n$ to the receiver. Till this point, we can formulate the SNR in the second hop as follows:
\begin{equation}\label{equ_gamma2}
\gamma_2 = \frac{\left|  {\bf x}^T D(\hat{\bf g}) {\bf y} + \sqrt{p_t}\hat{\bf f}_0^H {\bf w}_2 \right|^2}{1 +  ||D(\hat{\bf g}){\bf x}||^2 },
\end{equation}
where ${\bf x}= [x_1,x_2,\ldots, x_{|\mathcal{N}_a|}]^T$, ${\bf y}= [y_1,y_2,\ldots, y_{|\mathcal{N}_a|}]^T$ and $D(\hat{\bf g})$ denotes the diagonal matrix with the diagonal element given by $\hat{\bf g}= [\hat g_1,\hat g_2,\ldots, \hat g_{|\mathcal{N}_a|}]^T$.

Hence, the overall SNR at the receiver can be evaluated as $\gamma = \gamma_1 + \gamma_2$. We aim to maximize SNR in two hops by optimizing the HAP's beamforming strategies $({\bf w}_1,{\bf w}_2)$, as well as the relays' radio mode selection $b_n$ and operating parameters, including the power splitting factor $\rho_n$ and the complex reflection coefficient $\Gamma_n=|\Gamma_n|e^{j\theta_n}$:
\begin{subequations}\label{prob_bin}
\begin{align}
\max_{ {\bf w}_1, {\bf w}_2, b_n, {\rho_n}, \theta_n} ~&~ \gamma_1 + \gamma_2 \label{obj_sumsnr}\\
s.t.
~&~ ||{\bf w}_1 || \leq 1 \text{ and } ||{\bf w}_2|| \leq 1,\label{con_bvector}\\
~&~ b_n \in \{0,1\}, \quad \forall\,\, n\in\mathcal{N},\\
~&~ \theta_n \in [0, 2\pi] ,\quad \forall\,\, n\in\mathcal{N}_b.\label{con_gamma}\\
~&~ \rho_n \in(0,1), \quad \forall\,\, n\in\mathcal{N}_a, \label{con_rho}\\
~&~ p_n \leq    \eta \rho_n  p_t |\hat{\bf f}_n^H{\bf w}_1|^2, \quad \forall \,\, n\in\mathcal{N}_a.\label{con_power}
\end{align}
\end{subequations}
The constraints in~\eqref{con_bvector} denote the HAP's feasible beamforming vectors in two hops. We assume that the HAP's transmit power is fixed at $p_t$ while the beamforming vectors $({\bf w}_1,{\bf w}_2)$ can be adapted to maximize the throughput performance. Generaly ${\bf w}_2$ is not necessarily the same as ${\bf w}_1$ as the optimization of ${\bf w}_1$ has to take into account the data transmissions to both the relays and the receiver. The binary variable $b_n$ determines the division $(\mathcal{N}_a, \mathcal{N}_b)$ of the relays in different modes. The constraint in~\eqref{con_gamma} ensures that the phase offset of each passive relay in set $\mathcal{N}_b$ is fully controllable via load modulation~\cite{ieeenetwork}. The constraints in~\eqref{con_rho} and~\eqref{con_power} determine the active relays' transmit power in the second hop, which is upper bounded by the energy harvested from the HAP's signal beamforming in the first hop. Note that the power budget constraint~\eqref{con_power} is based on a linear energy harvesting model with a constant power conversion efficiency $\eta$, which may require an ideal circuit design for energy harvester. In practice, due to the nonlinearity of diode devices, the measurement results show a nonlinear energy harvesting model as proposed in~\cite{noneh}. In this case, we can consider a successive linear approximation of the nonlinear power budget constraint. Then our focus is the solution to the sub-problem in each iteration, which has a similar form as the problem in~\eqref{prob_bin}.

Besides power budget constraints for the active relays, the passive relays in set $\mathcal{N}_b$ are also wirelessly powered via energy harvesting. Let $p_c$ denote the constant power consumption of the passive relays. Then, we can formulate the passive relays' power budget as follows:
\[
p_c \leq (1-|\Gamma_n|^2) p_t \left( |\hat{\bf f}_n^H{\bf w}_1|^2 + |\hat{\bf f}_n^H{\bf w}_2|^2\right), \quad \forall \,\, n\in\mathcal{N}_b,
\]
where the scaling factor $(1-|\Gamma_n|^2)$ denotes the part of RF power harvested as energy. For simplicity, we assume that each passive relay sets the same reflection coefficient in both hops. The tuning of the reflection coefficient $\Gamma_n$ is subject to a limited range $[\Gamma_{\min},\Gamma_{\max}]$ due to the antenna's structural scattering effect~\cite{ieeenetwork}. The maximum magnitude $\Gamma_{\max}$ is typically less than one. With a small $p_c$, the passive relays' power budget constraints easily hold, and hence we omit it in the problem formulation.

\section{Performance Maximization with Hybrid Relay Communications}\label{sec_max}

It is clear that the optimization of the relays' mode selection $(\mathcal{N}_a, \mathcal{N}_b)$ is combinatorial and difficult to solve optimally. Even with fixed relay mode, the throughput maximization is still challenged by the couplings of multiple relays in different modes, which have very different tranmission capabilities and energy demands. In particular, the passive relays' operating parameters, e.g.,~the reflecting phase and magnitude, affect the active relays' channel conditions. A joint optimization is required to characterize their couplings and optimize all users' operating parameters simultaneously. However, the optimal solution is generally unavavilable. In the sequel, we propose to solve the throughput maximization problem in a decomposed manner. Firstly, assuming a fixed relay mode, we evaluate the enhanced channels and formulate the throughput maximization with only active relays, similar to that in~\cite{jc-iot}. Secondly, with the fixed beamforming strategy, we evaluate individual relays' energy status or performance gain. This motivates our algorithm design to update the relays' mode selection in an iterative manner.

\subsection{Relay Performance with Fixed Mode Selection}

Given a set $\mathcal{N}_b$ of the passive relays and their reflection coefficients $\Gamma_{n}$, the enhanced channels for active RF communications are given as in~\eqref{equ_channel_direct} and~\eqref{equ_channel_relay}. Then, we can formulate the throughput maximization problem with the set of active relays alone, which becomes a conventional two-hop relay optimization problem similar to that in~\cite{networkbf} and~\cite{jc-iot}. Our target is to maximize $\gamma$ by optimizing the HAP's beamforming $({\bf w}_1,{\bf w}_2)$ in two hops and the active relays' PS ratios $\boldsymbol{\rho}$, subject to the relays' power budget constraints:
\begin{subequations}\label{prob_bin_fixed}
\begin{align}
\max_{ {\bf w}_1, {\bf w}_2, \rho_n} ~&~ \gamma_1 + \gamma_2 \label{obj_bin_fixed}\\
s.t.
~&~ ||{\bf w}_1 || \leq 1 \text{ and } ||{\bf w}_2|| \leq 1, \\
~&~ \rho_n \in(0,1), \quad \forall\,\, n\in\mathcal{N}_a\\
~&~ p_n \leq    \eta \rho_n  p_t |\hat{\bf f}_n^H{\bf w}_1|^2, \quad \forall \,\, n\in\mathcal{N}_a,\label{con_powerbd}
\end{align}
\end{subequations}
Note that problem~\eqref{prob_bin_fixed} will achieve different performance when the passive relays set different phase offsets $\theta_{n}$. The phase optimization can follow the alternating optimization method. In particular, each passive relay initially sets a random phase offset $\theta_n$, based on which we can optimize $({\bf w}_1, {\bf w}_2)$ and $\rho_n$ by solving the problem~\eqref{prob_bin_fixed}. Given the solution to~\eqref{prob_bin_fixed}, we then turn to phase optimization sequentially for each passive relay, which will be detailed in Section~\ref{subsec_phase}.

\subsubsection{Lower Bounds on Relay Performance}
The throughput maximization~\eqref{prob_bin_fixed} is still challenging due to the non-convex coupling between different active relays in the objective~\eqref{obj_bin_fixed}. The HAP's beamforming strategy ${\bf w}_1$ is also coupled with the relays' PS ratio $\boldsymbol{\rho}$ in a non-convex form via the power budget constraint~\eqref{con_powerbd}. In the sequel, we provide a feasible lower bound on~\eqref{prob_bin_fixed}, which is achievable by designing the HAP's beamforming and relaying strategies.
\begin{proposition}\label{pro_convex}
A feasible lower bound on~\eqref{prob_bin_fixed} can be found by the convex reformulation as follows:
\begin{subequations}\label{prob_ps_convex}
\begin{align}
\max_{\bar{\bf W}_1, {\bf W}_1}~&~ p_t||\hat {\bf f}_0||^2 + p_t \hat{\bf f}_0^H {\bf W}_1 \hat{\bf f}_0 + p_t\sum_{n\in\mathcal{N}_a} s_{n,1} \label{obj_convex}\\
s.t. ~& \left[\begin{matrix}
            \kappa_n \psi_n - (1+  \psi_n) s_{n,1} & \sqrt{p_t}s_{n,1}\\
            \sqrt{p_t}s_{n,1}                 & 1
            \end{matrix}
            \right] \succeq 0, \,\, \forall n\in\mathcal{N}_a, \label{con_lmi}\\
~&~ \kappa_n \leq \hat{\bf f}_n^H {\bf W}_1\hat{\bf f}_n, \quad \forall n\in\mathcal{N}_a \label{con_power_ps1}\\
~&~ s_{n,1} = \hat{\bf f}_n^H {\bf W}_1\hat{\bf f}_n - \hat{\bf f}_n^H \bar {\bf W}_1\hat{\bf f}_n, \quad \forall n\in\mathcal{N}_a, \label{con_powermat_ps}\\
~&~ \bar{\bf W}_1\succeq {\bf 0} \text{ and } {\bf W}_1 \succeq {\bf 0}.
\end{align}
\end{subequations}
where $ \psi_n \triangleq  \eta p_t  |\hat g_n|^2 ||\hat {\bf f}_0||^2 $ is a constant. At optimum, the PS ratio of the relay-$n$ is given by $\rho_n = \frac{\hat{\bf f}_n^H\bar{{\bf W}}_1\hat{\bf f}_n}{\hat{\bf f}_n^H {\bf W}_1\hat{\bf f}_n}$ for $n\in\mathcal{N}_a$.
\end{proposition}
The proof of Proposition~\ref{pro_convex} follows a similar approach as that in~\cite{gsmwcnc19}, and thus we omit it here for conciseness. With the fixed relay mode, the channel information $\hat {\bf f}_0$ and $ \hat {\bf f}_n $ can be estimated by a training process. The objective function in~\eqref{obj_convex} then becomes linear and the constraints~\eqref{con_lmi}-\eqref{con_powermat_ps} define a set of linear matrix inequalities\footnote{The constraint~\eqref{con_powermat_ps} can be rewritten into two linear matrix inequalities.}. Hence, the resulting problem can be efficiently solved by semidefinite programming (SDP)~\cite{semi}. Once we find the optimal matrix solution $ {\bf W}_1$, we can retrieve the HAP's beamforming vector ${\bf w}_1$ by eigen-decomposition or Gaussian randomization method~\cite{rankluo}.

Though exact solution to~\eqref{prob_bin_fixed} is not available, problem~\eqref{prob_ps_convex} provides a lower bound on the SNR performance, which can serve as the performance metric for the relay's mode selection. It is clear that~\eqref{con_power_ps1} will hold with equality at optimum and thus we can verify the following property.
\begin{proposition}\label{pro_equality}
At optimum, the solution to~\eqref{prob_ps_convex} is given by $s_{n,1} = (\bar \rho_n \psi_n -1)/p_t$ , and the objective~\eqref{obj_convex} is given by
\begin{equation}\label{equ_newobj}
\gamma = p_t||\hat {\bf f}_0||^2 +  p_t|\hat{\bf f}_0^H {\bf w}_1|^2 + \sum_{n\in\mathcal{N}_a} \left( \frac{\eta\rho_n p_t}{1-\rho_n} |\hat g_n|^2||\hat {\bf f}_0||^2 -1 \right).
\end{equation}
\end{proposition}
The proof of Proposition~\ref{pro_equality} can be referred to~\cite{xiegc19}. It implies that the lower bound on SNR performance in~\eqref{prob_bin_fixed} can be evaluated directly from~\eqref{equ_newobj}. However, by an inspection on~\eqref{equ_newobj}, the SNR evaluation can be much less than the optimum of~\eqref{prob_bin_fixed} if the direct link $\hat {\bf f}_0$ is practically weak due to physical obstructions. To this end, we also provide another lower bound on problem~\eqref{prob_bin_fixed} by ignoring the direct link in the problem formulation. This corresponds to the case when the direct link is blocked or there is a long distance between the transceivers, e.g.,~\cite{liu2016wireless}. In this case, the objective in~\eqref{obj_bin_fixed} is lower bounded by $\gamma_1+\gamma_2 \geq \frac{\left|  {\bf x}^T D(\hat{\bf g}) {\bf y}  \right|^2}{1 +  ||D(\hat{\bf g}){\bf x}||^2 }$. Define $\hat x_n = x_n {\hat g}_n$ and then the lower bound on~\eqref{prob_bin_fixed} can be evaluated as follows:
\begin{subequations}\label{prob_lb}
\begin{align}
\max_{ \boldsymbol{\rho}, \hat{\bf x}, {\bf y}, {\bf w}_1 } ~&~  \left| \hat{\bf x}^T{\bf y}  \right|^2(1 + ||\hat{\bf x}||^2 )^{-1} \label{obj_lb}\\
s.t.
~&~ \hat x_n^2 \leq  \bar x_n(\rho_n,s_n) \triangleq\frac{ \eta \rho_n  p_t s_n^2 \hat g_n^2}{1+ (1-\rho_n)p_t s_n^2}  ,\label{con_x}\\
~&~ y_n^2 \leq  \bar y_n(\rho_n,s_n) \triangleq (1-\rho_n)p_t s_n^2, \label{con_y}\\
~&~ \rho_n \in(0,1), \quad \forall\,\, n\in\mathcal{N}_a\\
~&~ \hat{\bf x}\succeq {\bf 0}, {\bf y}\succeq {\bf 0},  \text{ and } ||{\bf w}_1 || \leq 1,
\end{align}
\end{subequations}
where $\hat {\bf x} \triangleq [\hat x_1, \hat x_2,\ldots, \hat x_{|\mathcal{N}_a|}]^T$ and we define $s_n^2 = |\hat {\bf f}_n^H {\bf w}_1|^2$ for notational convenience. The upper bounds on $\hat x_n^2$ and $y_n^2$ are defined as $\bar x_n(\rho_n,s_n)$ and $\bar y_n(\rho_n,s_n)$, respectively, which depend on the relays' PS ratio $\rho_n$ and the HAP's beamforming strategy ${\bf w}_1$. Note that $\hat{\bf x}$ and ${\bf y}$ are auxiliary decision variables in problem~\eqref{prob_lb}, coupled with the PS ratio $\boldsymbol{\rho}$ and the beamformer ${\bf w}_1$ via constraints in~\eqref{con_x}-\eqref{con_y}.

\subsubsection{Alternating Optimization Solution to~\eqref{prob_lb}}

To the best of our knowledge, there is no exact solution to the non-convex problem~\eqref{prob_lb}. Practically, it can be solved by the alternating optimization method that improves the objective~\eqref{obj_lb} in an iterative manner with guaranteed convergence. In particular, with fixed $\rho_n$ and ${\bf w}_1$, the auxiliary variables $\hat{\bf x}$ and ${\bf y}$ are subject to the fixed upper bounds $\bar x_n(\rho_n,s_n)$ and $\bar y_n(\rho_n,s_n)$, respectively. As such, problem~\eqref{prob_lb} can be viewed as the conventional network beamforming optimization problem with perfect channel information~\cite{networkbf}, which can be solved optimally in a closed form. However, given the feasible $\hat{\bf x}$ and ${\bf y}$, the optimization of $\rho_n$ and ${\bf w}_1$ is still very difficult due to their bilinear coupling in the constraints. This implies that we require further approximation within each iteration of the alternating optimization method. To proceed, we first exploit the structural properties of problem~\eqref{prob_lb} that shred some insight on the algorithm design.
\begin{proposition}\label{pro_property}
Problem~\eqref{prob_lb} has the following properties: (i) $\bar x_n(\rho_n, s_n)$ is increasing in both $\rho_n$ and $s_n$, (ii) $\bar y_n(\rho_n,s_n)$ is increasing in $s_n$ and decreasing in $\rho_n$, and (iii) the constraint~\eqref{con_y} holds with equality at the optimum of~\eqref{prob_lb}.
\end{proposition}
The proof of property (ii) in Proposition~\ref{pro_property} is straightforward by inspection. To verify property (i), we can simply rewrite the upper bound on $\hat x_n^2$ as follows:
\[
\bar x_n(\rho_n, s_n) = \frac{ \eta \rho_n  p_t s_n^2\hat g_n^2}{1+ (1-\rho_n)p_t s_n^2} = \frac{ \eta   p_t \hat g_n^2 }{(1/s_n^2 + p_t)/\rho_n - p_t },
\]
which is obviously increasing in $\rho_n$ and $s_n$. Then, we focus on the property (iii). Note that $y_n\geq 0$ only appears in the objective function. If~\eqref{con_y} holds with strict inequality at the optimum, we can simply improve the objective~\eqref{obj_lb} by increasing $y_n$ properly while keeping the other variables unchanged, which brings a contradition. Therefore, we can guarantee that $y_n = \sqrt{(1-\rho_n)p_t}s_n$ at the optimum of~\eqref{prob_lb}.

As both $\bar x_n(\rho_n,s_n)$ and $\bar y_n(\rho_n,s_n)$ are increasing functions of the variable $s_n^2$, we consider a replacement of $s_n$ by its smallest value $s_{\min}$. As such, we can further derive a lower bound on~\eqref{prob_lb} by the following problem:
\begin{subequations}\label{prob_lb2}
\begin{align}
\max_{ \boldsymbol{\rho}, \hat {\bf x}\succeq {\bf 0}, {\bf y}\succeq {\bf 0} } ~&~  \left| \hat{\bf x}^T{\bf y}  \right|^2(1 + ||\hat{\bf x}||^2 )^{-1} \label{obj_lb2}\\
s.t.
~&~ \hat x_n^2 \leq \bar x_n(\rho_n,s_{\min})  ,\label{con_x2}\\
~&~ y_n^2 \leq \bar y_n(\rho_n,s_{\min}) , \label{con_y2}\\
~&~  \rho_n \in(0,1), \quad \forall\,\, n\in\mathcal{N}_a.
\end{align}
\end{subequations}
Here $s^2_{\min}$ is given by $s^2_{\min} = \max_{\lVert{\bf w}_1\rVert \leq 1}\min_{n\in\mathcal{N}_a} |\hat {\bf f}_n^H {\bf w}_1|^2$, which can be easily reformulated into an SDP as follows:
\begin{subequations}\label{prob_smin}
\begin{align}
\max_{s_{\min}, {\bf W}_1} ~&~ s^2_{\min} \label{obj_smin}\\
s.t. ~&~ \hat {\bf f}_n^H {\bf W}_1 \hat {\bf f}_n \geq s^2_{\min}, \quad \forall\, n\in\mathcal{N}_a,\label{con1_smin}\\
~&~ {\bf W}_1 \succeq {\bf 0} \text{ and } \text{trace}({\bf W}_1) \leq 1.\label{con2_smin}
\end{align}
\end{subequations}
By solving problem~\eqref{prob_smin} with the interior-point algorithms~\cite{semi}, we can easily determine the HAP's beamformer ${\bf w}_1$ from the matrix solution ${\bf W}_1$ via Gaussian randomization.

Till this point, we can employ the alternating optimization method to solve~\eqref{prob_lb2}, which provides the lower bound on~\eqref{prob_lb} at the convergence. With a fixed PS ratio $\rho_n$, the feasible solution $(\hat{\bf x}, {\bf y})$ to~\eqref{prob_lb2} can be easily obtained by the network beamforming optimization in~\cite{networkbf}, and then we turn to update $\rho_n$ for each active relay based on the following property:
\begin{proposition}\label{pro_rho}
If the constraints in~\eqref{con_x2} hold with a strict inequality for some $n\in\mathcal{N}_a$ at optimum, e.g.,~$\hat x_n^2 < \bar x_n(\rho_n,s_{\min})$, we can further improve~\eqref{obj_lb2} by decreasing $\rho_n$.
\end{proposition}
The proof is straightforward by checking the properties in Proposition~\ref{pro_property}, which reveal that $\bar x_n(\rho_n, s_{\min})$ is increasing in $\rho_n$, while $y_n$ is decreasing in $\rho_n$. At the optimum of~\eqref{prob_lb2}, if $\hat x^2_n <\bar x_n(\rho_n, s_{\min})$ for some $n\in\mathcal{N}_a$, we can choose $\Delta_{n}>0$ such that $\bar x^2_n \leq \bar x_n(\rho_n - \Delta_{n} , s_{\min}) < \bar x_n(\rho_n, s_{\min}) $. Meanwhile, we have $y_n^2=\bar y_n(\rho_n,s_{\min}) < \bar y_n(\rho_n- \Delta_{n},s_{\min})$, which implies that the inequality constraint on $y_n$ can be relaxed. With the relaxed upper bounds on $\hat{\bf x}$ and ${\bf y}$, the network beamforming optimization in~\cite{networkbf} will produce a higher objective value.

All the above derivations lead to the iterative procedure in Algorithm~\ref{alg_ao}. The algorithm starts from a random initialization of $\rho_n$. The HAP's beamforming strategy ${\bf w}_1$ can be obtained by solving~\eqref{prob_smin} and fixed during the algorithm iteration. With the fixed $(\rho_n, {\bf w}_1)$, the upper bounds $(\bar{x}_n,\bar{ y}_n)$ are also fixed, and thus we can apply the network beamforming optimization to solve $(\hat{\bf x}, {\bf y})$ in problem~\eqref{prob_lb2}. According to Proposition~\ref{pro_rho}, the update the PS ratio $\rho_n$ can be based on the feasibility check of the inequality constraints in~\eqref{con_x2}-\eqref{con_y2}. In particular, for $n\in\mathcal{N}_a$, we search for the relay-$n$ with the largest constraint gap, defined as $G_n(\rho_n,s_{\min}) \triangleq \bar x_n(\rho_n, s_{\min}) - \hat x_n^2 $, and then we reduce $\rho_n$ properly by a small amount $\Delta_n$ such that
\[
\bar x_n(\rho_n-\Delta_n, s_{\min}) = \bar x_n(\rho_n , s_{\min}) - \beta G_n(\rho_n,s_{\min}),
\]
where $\beta\in (0,1)$ is a constant parameter. By solving above equation, we can determine $\Delta_n$ easily as follows:
\begin{equation}\label{equ_delta}
\Delta_n = \frac{\left(\frac{1}{p_ts_{\min}^2} +1\right)(\bar x_n(\rho_n , s_{\min}) - \beta G_n(\rho_n,s_{\min}))}{\eta + \bar x_n(\rho_n , s_{\min}) - \beta G_n(\rho_n,s_{\min})}
\end{equation}
Once $\rho_n$ is updated as shown in line 9 of Algorithm~\ref{alg_ao}, we turn to optimize $(\hat{\bf x},{\bf y})$ by network beamforming optimization~\cite{networkbf}.

\begin{algorithm}[t]
\caption{Alternating Optimization Solution to~\eqref{prob_lb}}\label{alg_ao}
\begin{algorithmic}[1]
\State Initialize $\rho_n$ for $n\in\mathcal{N}_a$, set ${\bf w}_1$ by solving~\eqref{prob_smin}
\State ${\gamma^{(0)}}\leftarrow 0$, $t\leftarrow 1$, $\gamma^{(t)}\leftarrow p_t|\hat{\bf f}_0^H {\bf w}_1|^2$, $\epsilon \leftarrow 10^{-5}$, $\beta \leftarrow 1/2$
\State {\bf while} $ \lvert \gamma^{(t)} - \gamma^{(t-1)} \rvert > \epsilon $
\State \hspace{5mm} {$t\leftarrow t+1$}
\State \hspace{5mm} Update $\bar x_n(\rho_n,s_{\min})$ and $\bar y_n(\rho_n,s_{\min})$
\State \hspace{5mm} Update $\gamma^{(t)}$ and $(\hat{\bf x}, {\bf y})$ by solving problem~\eqref{prob_lb2}
\State \hspace{5mm} Evaluate $G_n(\rho_n,s_{\min})$ for $n\in\mathcal{N}_a$
\State \hspace{5mm} $m\leftarrow \arg\max_{n\in\mathcal{N}_a}G_n(\rho_n,s_{\min})$
\State \hspace{5mm} $\rho_{m} \leftarrow \rho_{m} - \Delta_{m}$, where $\Delta_{m}$ is given by~\eqref{equ_delta}
\State {\bf end while}
\end{algorithmic}
\end{algorithm}

Moreover, we can show that the lower bound derived by Algorithm~\ref{alg_ao} is also applicable to a single-antenna case. In parituclar, when the HAP has one single antenna, the enhanced channels $\hat f_n$ from the HAP to different relays now become complex variables instead of vectors. The power budget constraints in~\eqref{equ_lineareh} are degenerated to $p_n \leq \eta\rho_{n}p_t|f_n|^2$. Each active relay's power amplifying coefficient can be similarly defined as $ x_n =\left(\frac{p_n}{1+y_n^2}\right)^{1/2}$ where $y_n\triangleq\sqrt{(1-\rho_n)p_t} f_n$. As such, the performance maximization problem is given as follows:
\begin{subequations}\label{pro_singel_ps}
\begin{align}
\max_{ \boldsymbol{\rho}, \hat{\bf x}, {\bf y} } ~&~  \left| \hat{\bf x}^T{\bf y}  \right|^2(1 + ||\hat{\bf x}||^2 )^{-1} \label{obj_sumsnr}\\
s.t.
~&~ \hat x_n^2 \leq  \bar x_n(\rho_n, |f_n|) = \frac{ \eta \rho_n  p_t |f_n|^2 \hat g_n^2}{1+ (1-\rho_n)p_t |f_n|^2}  ,\label{con_x_sin}\\
~&~ y_n^2 \leq  \bar y_n(\rho_n, |f_n|) =  (1-\rho_n)p_t |f_n|^2, \label{con_y_sin}\\
~&~ \hat{\bf x}\succeq {\bf 0}, {\bf y}\succeq {\bf 0},  \text{ and } \rho_n \in(0,1), \quad \forall\,\, n\in\mathcal{N}_a.
\end{align}
\end{subequations}
By simply setting $s_{n}^2 = |\hat f_n|^2$, problem~\eqref{pro_singel_ps} has the same form as that in~\eqref{prob_lb2} and hence it is solvable by Algorithm~\ref{alg_ao}.

\subsection{Iterative Algorithm for Relay Mode Selection}\label{subsec_phase}

The previous analysis provides the SNR evaluation with fixed relay mode. This can serve as the performance metric to update each relay's mode selection. The basic idea of an iterative algorithm for the relays' mode selection is to start from the special case with all active relays and then update the relay mode one by one depending on the relay's performance gain. For simplicity, we allow the mode switch of a single relay in each iteration. Hence, the number of iterations will be linearly proportional to the number of relays and the main computational complexity lies in the SNR evaluation given the division $(\mathcal{N}_a, \mathcal{N}_b)$ within each iteration. Such an iterative process continues until no further improvement can be achieved by changing the relays' mode.

\begin{algorithm}[t]
\caption{Phase Optimization for SNR Performance}\label{alg_max_snr}
\begin{algorithmic}[1]
\State Initialize $({\bf w}_1, \rho_n,\Gamma_{n})$ , $\epsilon\leftarrow 10^{-5}$, $M\leftarrow 20$
\State SNR$_n^{(0)}\leftarrow 0$, $t\leftarrow 1$, SNR$_n^{(t)}\leftarrow p_t||\hat {\bf f}_0||^2+  p_t|\hat{\bf f}_0^H {\bf w}_1|^2 $

\State {\bf while} $|$SNR$_n^{(t)} -$ SNR$_n^{(t-1)}|>\epsilon$
\State \hspace{5mm} $t\leftarrow t+1$
\State \hspace{5mm} Fix the solution $({\bf w}_1, {\bf \rho}_n)$
\State \hspace{5mm} \emph{Phase optimization: }
\State \hspace{5mm} $\theta_n\leftarrow \arg_{\gamma\in \Theta_d}\max$ SNR$_n(\gamma)$ by solving~\eqref{opt_theta}
\State \hspace{5mm} SNR$_n^{(t)}\leftarrow$ SNR$_n(\theta_n)$
\State \hspace{5mm} \emph{Beamforming optimization: }
\State \hspace{5mm} Fix the phase $\theta_n$ and update channels by~\eqref{equ_channel_direct}-\eqref{equ_channel_relay}
\State \hspace{5mm} Update (${\bf w}_1$, ${\bf \rho}_n$) by solving~\eqref{prob_ps_convex} or~\eqref{prob_lb}
\State {\bf end while}
\State {\bf return} Maximum SNR$_n$ and the phase $\theta_n$
\end{algorithmic}
\end{algorithm}

\subsubsection{Evaluation of SNR Performance}  Considering different channel conditions in the direct link, the SNR evaluation can be performed by solving either the SDP in~\eqref{prob_ps_convex} or the non-convex problem in~\eqref{prob_lb} by Algorithm~\ref{alg_ao}. With the fixed $({\bf w}_1,\boldsymbol{\rho})$, we can evaluate the SNR performance of each relay when it is in the passive mode. After iterating over all relays, we can switch the relay with the maximum SNR performance to the passive mode. It is obvious that the SNR evaluations in~\eqref{prob_ps_convex} and~\eqref{prob_lb} both rely on the passive relay's complex reflection coefficient $\Gamma_n = |\Gamma_n|e^{j\theta_n}$, which is critical for the channel enhancement in~\eqref{equ_channel_direct} and~\eqref{equ_channel_relay} and thus effects the SNR evaluation. To maximize the SNR performance, the passive relays can simply set the magnitude of reflection $|\Gamma_n|$ to its maximum $\Gamma_{\max}$. However the complex phase $\theta_n$ is more difficult to optimize due to its couplings cross different relays. The dependence of different channels makes it difficult to enhance all active relays' channels simultaneously.

The optimization of phase $\theta_n$ can follow the alternating optimization method. In particular, we can firstly optimize the beamforming strategy $({\bf w}_1, {\bf w}_2)$ and the relays' PS ratios $\boldsymbol{\rho}$ to maximize the SNR performance. After that, we fix $({\bf w}_1, {\bf w}_2)$ and $\boldsymbol{\rho}$ and then turn to optimize the passive relay's reflecting phase $\theta_n\in [0,\pi]$ to further improve the SNR performance. Considering the lower bound in~\eqref{equ_newobj}, the maximum SNR via phase optimization can be approximated as follows:
\begin{align}\label{prob_phase}
\max_{\theta_n\in [0,\pi]} \,&\, p_t||\hat {\bf f}_0||^2 +  p_t|\hat{\bf f}_0^H {\bf w}_1|^2 + \eta p_t ||\hat {\bf f}_0||^2 g_t,
\end{align}
where we simply replace $\sum_{n\in\mathcal{N}_a} \frac{ \rho_n }{1-\rho_n} | \hat g_n|^2$ in~\eqref{equ_newobj} by a known approximation $g_t$. This approximation stems from the observation that different active relays are in general spatially distributed with indepedent channel conditions. As such, the optimization of complex reflection coefficient $\theta_n$ of one single passive relay has very limited capability to enhance all the active relays' forwarding channels $\hat g_n$ simultaneously. For example, the forwarding channel $\hat g_m$ of an active relay-$m$ can be enhanced by the passive relay-$n$ with the reflection coefficient $\Gamma_n$ (i.e.,~$|\hat g_m|^2>|g_m|^2$), while another channel $\hat g_k$ may become weakened (i.e.,~$|\hat g_k|^2<|g_k|^2$) due to the indepedence among different relays. Hence, we simply view the term $\sum_{n\in\mathcal{N}_a} \frac{ \rho_n }{1-\rho_n} | \hat g_n|^2$ as a constant approximated by $g_t \triangleq \sum_{n\in\mathcal{N}_a} \frac{ \rho_n }{1-\rho_n} | g_n|^2$. By this approximation, problem~\eqref{prob_phase} can be further rewritten as follows:
\begin{align}\label{opt_theta}
\max_{\theta_n\in [0,\pi]} & \hat {\bf f}_0^H ( (1+\eta g_t) {\bf I} + {\bf w}_1{\bf w}_1^H ) \hat{\bf f}_0.
\end{align}
Let ${\bf A} = (1+\eta g_t) {\bf I} + {\bf w}_1{\bf w}_1^H \succeq {\bf 0}$ be a known matrix coefficient were ${\bf I}$ denotes the identity matrix. It is clear that problem~\eqref{opt_theta} aims to maximize the compound channel gain $\hat {\bf f}_0^H {\bf A}\hat{\bf f}_0$ given the HAP's beamforming strategy ${\bf w}_1$, which can be easily solved by one-dimension search algorithm.

Let SNR$_n(\theta_n)$ denote the objective in~\eqref{opt_theta} when the relay-$n$ is selected as the passive relay with the reflecting phase $\theta_n$. Assuming a fixed reflection magnitude $\Gamma_{\max}$, the enhanced channel $\hat {\bf f}_0 $ in~\eqref{equ_channel_direct} can be simplified as $\hat{\bf f}_0  = {\bf f}_0+ e^{j\theta_n}\Gamma_{\max}g_n{\bf f}_n$. The optimal phase $\theta_n^*$ can be simply obtained by a one-dimension search method. In particular, we can quantize the continuous feasible region $[0,\pi]$ into a finite discrete set $\Theta_d\triangleq \{0, \pi/M, 2\pi/M,\ldots,\pi\}$, where $M$ denotes the size of $\Theta_d$. Then, we can devise a one-dimension search algorithm to optimize the phase parameter $\theta_n^*$ that maximizes SNR$_n(\theta_n)$. The detailed solution procedure is presented in Algorithm~\ref{alg_max_snr}, which alternates between the HAP's beamforming optimization and the relays' phase optimization by solving two sub-problems in~\eqref{prob_ps_convex} and~\eqref{opt_theta}, respectively. The iteration terminates when SNR$_n(\theta_n)$ can not be improved anymore.

\begin{algorithm}[t]
\caption{Max-SNR based Relay Mode Selection}\label{alg_gain}
\begin{algorithmic}[1]
\State Initialize SNR$_0$ by solving (\ref{prob_ps_convex}) with all active relays
\State $\gamma^{(0)}\leftarrow 0$, $t\leftarrow 1$, $\gamma^{(t)}\leftarrow$ SNR$_0$
\State $\mathcal{N}_a \leftarrow \mathcal{N}$, $\mathcal{N}_b \leftarrow \emptyset$, $\epsilon\leftarrow 10^{-5}$
\State {\bf while} $ |\gamma^{(t)} > \gamma^{(t-1)} |>\epsilon$
\State \hspace{5mm} $t\leftarrow t+1$, $\gamma^{(t)}\leftarrow \gamma^{(t-1)}$
\State \hspace{5mm} {\bf for} $n\in\mathcal{N}_a$
\State \hspace{9mm} Update channels $\hat{\bf f}_0$ and $\hat{\bf f}_n$ by~\eqref{equ_channel_direct} and~\eqref{equ_channel_relay}
\State \hspace{9mm} Update SNR$_n(\theta_n^*)$ and $({\bf w}_1, {\bf \rho}_n)$ by Algorithm~\ref{alg_max_snr}
\State \hspace{5mm} {\bf end for}
\State \hspace{5mm} {\bf if} $ \max_{n\in\mathcal{N}}$ SNR$_n(\theta_n^*) > \gamma^{(t-1)}$
\State \hspace{9mm} $\gamma^{(t)}\leftarrow \max_{n\in\mathcal{N}}$ SNR$_n(\theta_n^*)$
\State \hspace{9mm} $n^*\leftarrow \arg\max_{n\in\mathcal{N}}$ SNR$_n(\theta_n^*)$
\State \hspace{9mm} $\mathcal{N}_b \leftarrow \mathcal{N}_b \cup\{n^*\}$
\State \hspace{9mm} $\mathcal{N}_a \leftarrow \mathcal{N}_a - \{n^*\}$
\State \hspace{5mm} {\bf end if}
\State {\bf end while}
\State {\bf return} ${\bf w}_1$, $(\mathcal{N}_a, \mathcal{N}_b)$, $\{\rho_n\}_{n\in\mathcal{N}_a}$, and $\{\theta_n\}_{n\in\mathcal{N}_b}$
\end{algorithmic}
\end{algorithm}

\subsubsection{Simplified Schemes for Relay Mode Selection}
The convergent value of SNR$_n(\theta_n)$ in Algorithm~\ref{alg_max_snr} can be used as a performance metric to evaluate the relay performance when the relay-$n$ is selected as the passive relay. We denote this SNR-based performance metric as the Max-SNR scheme. However, we note that the SNR evaluation in~\eqref{prob_ps_convex} or~\eqref{prob_lb} is quite complicated as it requires to solve optimization problems in an iterative procedure. In this part, besides the Max-SNR scheme, we also devise a set of simplified performance metrics based on the problem structure and the corresponding iterative algorithms to update the relays' mode selections.
\begin{itemize}
  \item Max-Direct-Rate (Max-DR): With a strong direct link from the HAP to the receiver, the data rate contributed by the direct links becomes significant. The Max-DR scheme aims to maximize the data rate via the direct links, which is given by $p_t||\hat {\bf f}_0||^2 +  p_t|\hat{\bf f}_0^H {\bf w}_1|^2$. The phase optimization follows a similar approach as that in~\eqref{opt_theta} while the beamforming optimization can be significantly simplified comparing to that of the SDP~\eqref{prob_ps_convex}.
  \item Max-Relay-Rate (Max-RR): In contrast to the Max-DR scheme, the Max-RR scheme picks the passive relay which acheives the maximum relay performance without direct links by solving problem~\eqref{prob_lb2}, which avoids the beamforming optimization in every iteration and thus has significantly reduced complexity than that of~\eqref{prob_ps_convex}.
  \item Max-Direct-Gain (Max-DG): The Max-DG scheme further simplifies the optimization in the Max-DR scheme, by focusing on the maximization of the channel gain $|\hat {\bf f}_0|^2$ instead of the data rate, which can be separated from the beamforming optimization. Hence, the Max-DG scheme avoids the iterative procedure, making it more efficient than the alternating optimization method for solving~\eqref{prob_lb2}.
  \item Min-RF-Energy (Min-RF): This scheme stems from the intuition that the passive radios are more energy efficient than the active radios. Hence, the relay with low energy supply may prefer to operate in the passive mode. Given the beamforming in~\eqref{prob_ps_convex}, we can sort the active relays by the RF power at their antennas, i.e., $p_n = \eta\rho_n  p_t |\hat{\bf f}_n^H{\bf w}_1|^2$. The Min-RF scheme switchs the active relay with the minimum RF power into the passive mode. 
\end{itemize}

The complete procedure for relay mode selection based on the Max-SNR metric is presented in Algorithm~\ref{alg_gain}. It is initialized with a conventional relay network when all relays are in the active mode. The initial SNR performance SNR$_0$ serves as the baseline for performance improvement. In lines 6-9 of Algorithm~\ref{alg_gain}, we sequentially select one node from the active relay set $\mathcal{N}_a$ and then evaluate the maximum SNR performance SNR$_n(\theta_n^*)$ when it is turned into the passive mode. If the selected relay can improve the overall SNR performance, we will remove it from $\mathcal{N}_a$ and add it to the passive relay set $\mathcal{N}_b$, as shown in lines 10-15 of Algorithm~\ref{alg_gain}. Note that the performance metric for relay mode selection in line 8 of Algorithm~\ref{alg_gain} can be changed to the Max-DR, Max-RR, Max-DG, and the Min-RF schemes. A performance comparison of different schemes will be presented in Section~\ref{sec_sim}.

The computational complexity of Algorithm~\ref{alg_gain} depends on the number of iterations in the outer loop and that of Algorithm~\ref{alg_max_snr} in the inner loop. As we only select one passive relay at each iteration, the total number of outer-loop iterations in Algorithm~\ref{alg_gain} will be in the order of $N$. Within each iteration of Algorithm~\ref{alg_gain}, Algorithm~\ref{alg_max_snr} is called to evaluate the SNR performance. Our numerical results show that the alternating optimization method in Algorithm~\ref{alg_max_snr} can converge quickly in a few iterations. However, within each iteration we require to solve the SDP~\eqref{prob_ps_convex}, which determines the computational complexity of Algorithm~\ref{alg_max_snr}. Given the size of an SDP problem, the computational complexity of~\eqref{prob_ps_convex} can be easily characterized by the analytical work in~\cite{rankluo}, which is a polynomial relating to the sizes of matrix variables and the number of constraints.

\section{Numerical Results}\label{sec_sim}

\begin{figure}[t]
\centering
\includegraphics[width=\singlesize\textwidth]{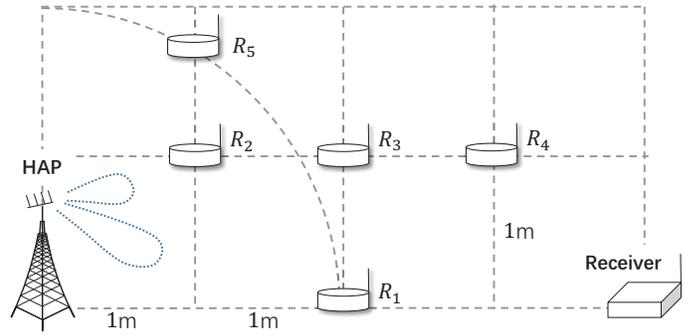}
\caption{Network topology of hybrid relay communications.}\label{fig_topo}
\end{figure}

In the simulation, we consider the multi-antenna HAP with $K=3$ antennas and totally $N=5$ dual-mode energy harvesting relays assisting the information transmission from the HAP to the receiver. To illustrate the radio mode switching of individual relays, we focus on a specific network topology as shown in Fig.~\ref{fig_topo}, where the distance in meters between the HAP and the receiver is denoted as $d_0=4$. The other relays are spatially distributed between the HAP and the receiver. The location of each relay is also depicted in Fig.~\ref{fig_topo}. The conclusions drawn based on a fixed topology may help us understand the connections between channel conditions and the relays' radio mode selections. Throughout the simulations we have the following parameter settings unless otherwise stated. The noise power density is $-90$~dBm and the bandwidth is $100$~kHz. The HAP's transmit power $p_t$ in milliwatts can be tuned from $10$ to $90$. The efficiency for energy harvesting is set to $\eta = 0.5$. The complex modulus of each passive relay's reflection coefficient can be fixed at $\Gamma_{\max}= 0.5$. We adopt the log-distance propagation model $L=L_0+10\alpha\log_{10}(d/d_0)$, where the loss exponent is $\alpha=2$ and pass loss in dB at the reference distance $d_0=1$~m is $L_0=30$. The antenna gain between the transceiver is set to $15$~dB, e.g., the HAP can use the directional Antenna ACD24W08 with 12 dBi gain and the user devices use commercial monopole antenna with 3.0 dBi gain. The phase shifts of complex channels are randomly generated. Similar parameter setting can be referred to~\cite{jc-iot}.

\subsection{Motivation for Hybrid Relay Communications}
Intuitively, the relays' radio mode selection depends on individual relays' channel and energy conditions, given different energy demands and transmission capabilities in two radio modes. To verify the performance gain in hybrid relay communications, we consider a three-node relay model with one relay inbetween the HAP and the receiver. By varying the channel conditions in two hops, we examine the relay's optimal radio mode that maximizes the relay performance.
\begin{figure}[t]
	\centering
	\includegraphics[width=\singlesize\textwidth]{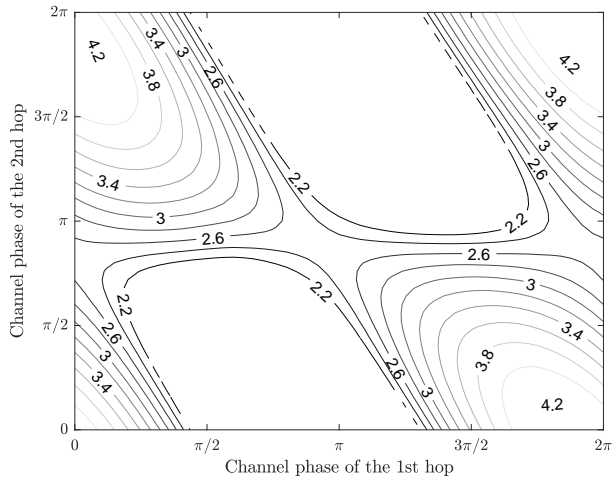}
	\caption{The relay's mode selection in a three-node relay model, where one relay locates inbetween the HAP and the receiver. The HAP's transmit power is fixed at $p_0=10$~mW. The distance $d_0$ from the HAP to the receiver increases from 2 to 5 meters.}\label{fig_mode_switch}
\end{figure}

\begin{figure}[t]
	\centering
	\includegraphics[width=\singlesize\textwidth]{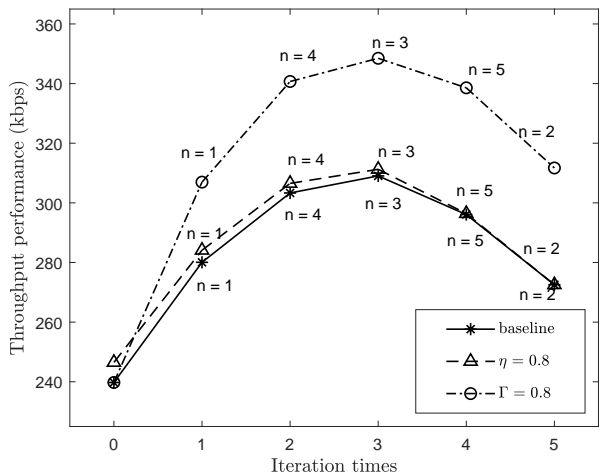}
	\caption{The overall relay performance with a different set of passive relays: 5 active relays are sequentially turned into the passive mode. In baseline case, we have the default setting $(\eta=0.5, \Gamma=0.5)$.}\label{fig_allpassive}
\end{figure}

Specifically, we fix the channel condition for the first hop between the HAP and the relay, and then vary the phase shift of the complex channel for the second hop from the relay to the receiver. With each fixed channel condition, we evaluate the relay performance with the single relay either in the passive or the active mode. The optimal relay mode is then selected to achieve the maximum relay performance. The numerical result demonstrates that the relay will switch to the passive mode when the phase shifts of the complex channel are set as $4\pi/5$ and $9\pi/5$. This shows some periodicity and verifies the importance of channel conditions on the relay's mode selection. It also implies that the passive relay's phase shift induced by backscattering has to be optimized to match the complex channel conditions. Furthermore, we fix the channel's phases in two hops, and then vary the transmission distance between the HAP and the receiver. Generally, a larger distance implies a higher attenuation to the signal propogation and thus makes the channels worse off. Similarly, we examine the relay performances in two modes and choose the optimal relay mode to maximize the overall throughput. As shown in Fig.~\ref{fig_mode_switch}, we plot the point of relay's mode switching when we set different transmission distances $d_0\in(2,5)$, i.e., the receiver moves far away from the HAP. Each point on the curve in Fig.~\ref{fig_mode_switch} indicates the relay's mode switch from the active to the passive mode. The number attached to each curve represents the distance between the HAP and the receiver at which the relay's mode switching happens. It is obvious that the relay's mode selection is also strongly coupled with the channel gains. In particular, the relay tends to work in the passive mode when the channels become worse off as the distance $d_0$ becomes large. We also note that there exists some cases in which the relay is always operating in the active mode as the distance $d_0$ varies.

To verify how passive relays affect the overall throughput performance, we allow all relays in Fig.~\ref{fig_topo} to switch into the passive mode one by one, and then we evaluate the throughput performance in each case. Given the relays' radio modes, we optimize the HAP's beamforming strategy and the relays' operating parameters to maximize the throughput performance. As shown in Fig.~\ref{fig_allpassive}, when we sequentially set the relay-$1$, relay-$4$, and relay-$3$ into the passive mode, the overall throughput performance can be gradually increased. However, the overall throughput becomes worse off when the relay-$2$ and relay-$5$ are further turned into the passive mode. The maximum throughput is achieved when both the passive and the active relays collaborate in the relay transmission. In particular, we have $\mathcal{N}_a = \{2,5\}$ and $\mathcal{N}_b = \{1,3,4\}$ for the network in Fig.~\ref{fig_topo}. This observation clearly verifies that the proposed hybrid relay communications can potentially outperform the conventional relay communications with all relays in the same radio mode.

\begin{figure}[t]
	\centering
	\includegraphics[width=\singlesize\textwidth]{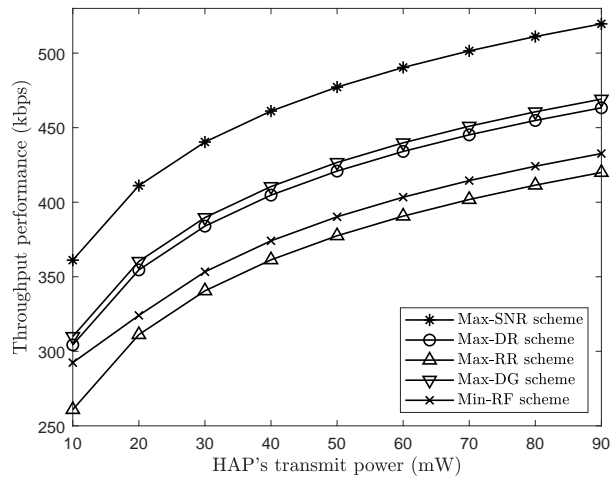}
	\caption{Comparison of different mode selection algorithms.}\label{fig_alg_compare}
\end{figure}

\subsection{Comparison of Different Mode Selection Algorithms}

For multiple relays in the system, the optimization of individual relays' mode selections becomes combinatorial and problematic. Along with the Max-SNR metric in Algorithm~\ref{alg_gain}, we have also proposed a set of heuristic algorithms with reduced complexity. The Max-SNR scheme evaluates a lower bound of the relay performance and relies on an iterative procedure to improve the relay performance by updating the relays' mode selection in each iteration. A few heuristic algorithms are also proposed to simplify the evaluation of the overall SNR performance. The Max-DR scheme focuses on the throughput maximization of the direct link from the HAP to the receiver, while the Max-RR scheme maximizes the relay performance without the direct links. The Max-DG scheme avoids the iterative procedure by simply maximizing the channel gain of the direct links. In the last, the Min-RF scheme avoids the trouble in solving optimization problems and updates the relays' mode selections based on an ordering of their RF energy at the antennas.

Figure~\ref{fig_alg_compare} shows the overall throughput performance by using different algorithms for the relays' mode selection. It is clear that the Max-SNR scheme significantly outperforms the other schemes, however with the cost of a higher computational complexity. Note that both of the Max-DR and the Max-RR schemes are relying on simplified approximations of the overall SNR performance. In Fig.~\ref{fig_alg_compare}, we observe that the Max-DR schemes achieves higher performance than that of the Max-RR scheme, which implies that the direct links contribute significantly to the overall relay performance in the evaluated network topology. An interesting observation is that the most simple Max-DG scheme even outperforms that of the Max-DR scheme, which requires an iterative procedure to jointly optimize the reflecting phase and the HAP's beamforming strategy. Therefore, we can conclude that the Max-DG scheme is practically good in terms of throughput performance and computational complexity. The Max-SNR scheme can be viewed as a performance benchmark. In the sequel, we focus on the Max-SNR scheme and verify its performance dynamics with respect to different parameters.

\subsection{Throughput Dyanmics in the Max-SNR Algorithm}

\begin{figure}[t]
	\centering
	\includegraphics[width=\singlesize\textwidth]{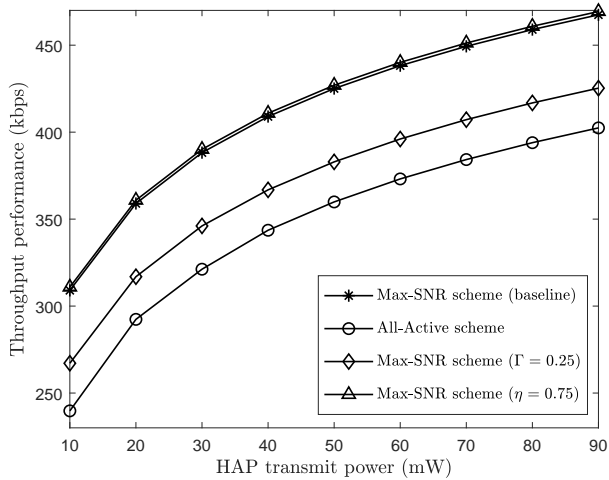}
	\caption{Throughput performance of the Max-SNR algorithm with different parameters.}\label{fig_maxsnr}
\end{figure}

In Fig.~\ref{fig_maxsnr}, we evaluate the throughput performance of the Max-SNR scheme with different transmit power at the HAP. For comparison, we also show the relay performance when all relays are working in the active mode, which is denoted as the All-Active scheme in Fig.~\ref{fig_maxsnr}. In the baseline algorithm, we set $\Gamma = 0.5$ and $\eta = 0.5$. These parameters are also varied to examine the performance of the Max-SNR algorithm when the relays have different capabilities in signal reflection and energy harvesting. The most straightforward observation is that the Max-SNR algorithm achieves the highest throughput compared to the All-Active scheme, which coincides with our observation in Fig.~\ref{fig_allpassive} and corroborates our motivation to optimize the relays' mode selection in hybrid relay communications. In particular, with 50 mW transmit power at the HAP, the Max-SNR algorithm achieves nearly 20 $\%$ throughput improvement comparing to that of the conventional All-Active scheme. Moreover, we observe that the overall relay performance barely increases when the relays can harvest RF energy more efficiently with a larger value $\eta=0.75$ comparing to the baseline $\eta=0.5$. This can be understood by the extremely low power consumption of the passive relays. Even with a smaller energy harvesting efficiency, the RF power can be still sufficient to power the passive relays' operations. As such, the overall relay performance becomes insensitive to the relays' energy harvesting, however it can be very sensitive to the passive relays' capabilities of signal reflections. As demonstrated in Fig.~\ref{fig_maxsnr}, the throughput is decreased significantly when we set a smaller value $\Gamma = 0.25$ comparing to the baseline $\Gamma = 0.5$.

\begin{figure}[t]
	\centering
	\includegraphics[width=\singlesize\textwidth]{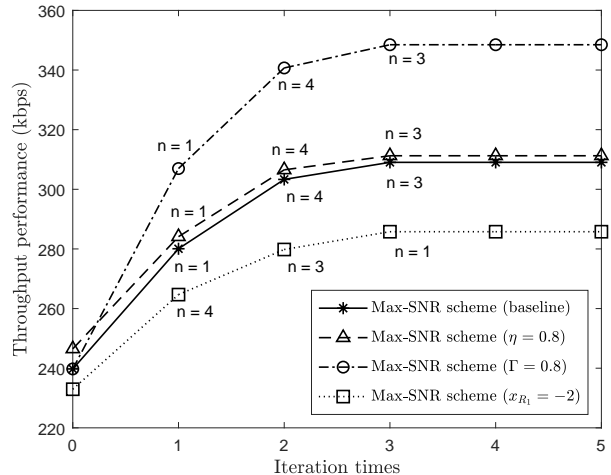}
	\caption{Throughput dynamics in the Max-SNR algorithm: The relay's mode selection improves the overall throughput in each iteration.}\label{fig_outerloop}
\end{figure}

\begin{figure}[t]
	\centering
	\includegraphics[width=\singlesize\textwidth]{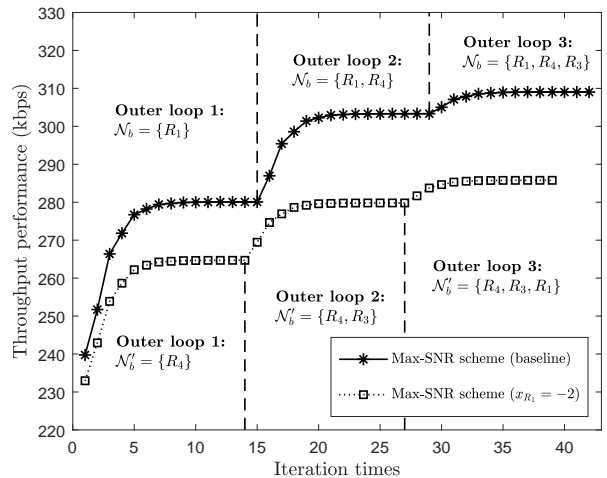}
	\caption{Throughput dynamics in the Max-SNR algorithm. The update of relay mode selection happens in the outer loop. Within each outer loop, the algorithm optimizes $({\bf w}_1, \boldsymbol{\rho})$ in an alternating manner.}\label{fig_innerloop}
\end{figure}
 
In this part, we intend to show how each relay changes its radio mode as the algorithm iterates. Fig.~\ref{fig_outerloop} shows the throughput dyanmics of the Max-SNR based Algorithm~\ref{alg_gain} in each iteration. We also show the index of the passive relay on the throughput curve when it is switched into the passive mode. As illustrated in Fig.~\ref{fig_outerloop}, the Max-SNR algorithm initializes all relays in the active mode. As the algorithm iterates, the relay-$1$, relay-$4$, and relay-$3$ are sequentially switched to the passive mode to improve the overall relay performance. The throughput dynamics in the inner loop of Algorithm~\ref{alg_gain} are detailed in Fig.~\ref{fig_innerloop}. Each outer loop implies the update of one relay's radio mode, while each inner loop indicates an iteration in the alternating optimization of ${\bf w}_1$ and $\boldsymbol{\rho}$. With the fixed relay mode, we can observe that the throughput is improved in each iteration of the alternating optimization method. Algorithm~\ref{alg_gain} terminates when the throughput performance cannot be further improved by changing the relays' radio mode. Note that the relay-$1$, relay-$4$, and relay-$3$ are generally far away from the HAP while closer to the receiver, as observed from the network topology in Fig.~\ref{fig_topo}. Hence, the results in Fig.~\ref{fig_outerloop} imply that the relays with worse channel conditions in the first hop may prefer to operate in the passive mode. With different parameters $\Gamma = 0.8$ and $\eta = 0.3$, the results in Fig.~\ref{fig_outerloop} verify that the throughput performance is insensitive to the relays' capabilities of energy harvesting but more sensitive to the capabilities of signal reflections. Fig.~\ref{fig_outerloop} also shows the throughput dynamics when the relay-$1$ changes its location (i.e.,~the $x$-coordinate of the relay-$1$ is shifted to $-2$), which implies the change of the relays' channel conditions and thus affects the sequence of radio mode switching in Algorithm~\ref{alg_gain}.

\begin{figure}[t]
	\centering
	\includegraphics[width=\singlesize\textwidth]{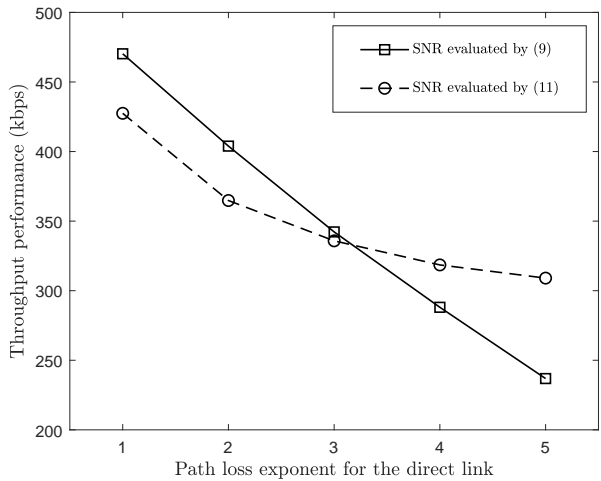}
	\caption{The throughput performance achieved by the Max-SNR algorithm, based on two different lower bounds for SNR evaluation.}\label{fig_twolb}
\end{figure}

The relay mode selection in Algorithm~\ref{alg_gain} is based on the SNR evaluation, which inlcudes the contributions from both the direct links and the relay channels. However, the SNR evaluation by problem~\eqref{prob_ps_convex} can be poor when the direct links are practically worse off, which will lead to the underestimated throughput performance in Algorithm~\ref{alg_gain}. To avoid this discrepancy, we also propose another performance lower bound in~\eqref{prob_lb} especially for the case with weak direct links. Based on the new lower bound, Algorithm~\ref{alg_gain} can be correspondingly enhanced to guide the search for the optimal set of passive relays. Fig.~\ref{fig_twolb} shows the comparison of throughput performance when we use two different lower bounds to evaluate the SNR performance in Algorithm~\ref{alg_gain}. A higher path loss exponent implies that the direct link from the HAP to the receiver becomes worse off. We observe that the SNR evaluation based on~\eqref{prob_ps_convex} performs much better than that based on~\eqref{prob_lb} when the direct link is strong with a small path loss exponent. As expected, this tendency becomes reversed when the direct link becomes severely antunuated with a large path loss exponent.

\section{Conclusions}\label{sec_con}

In this paper, we have introduced the novel concept of hybrid relay communications involving both the active and passive relays, and then presented a throughput maximization problem by jointly optimizing the HAP's beamforming, the relays' mode selection and operating parameters. Though the throughput maximization problem is non-convex, we have provided two lower bounds to evaluate the SNR performance under different channel conditions, which further serve as the performance metric for updating the relays' mode selection in an iterative manner. Besides the SNR-based performance metric, we have also devised a few heuristic performance metrics for the relays' mode selection with much reduced computational complexity. Simulation results have verified the advantageous of using hybrid relay communications, and demonstrated that a significant performance gain can be achieved by the proposed algorithms comparing to the conventional relay communications with all active relays.

\bibliographystyle{IEEEtran}

\bibliography{texreference}

\end{document}